# THE WAVEFRONT CONTROL SYSTEM FOR THE NATIONAL IGNITION FACILITY

Lewis Van Atta, Mark Perez, Richard Zacharias, and William Rivera
Lawrence Livermore National Laboratory, Livermore, CA 94550, USA

## Abstract

The National Ignition Facility (NIF) requires that pulses from each of the 192 laser beams be positioned on target with an accuracy of 50 μm rms. Beam quality must be sufficient to focus a total of 1.8 MJ of 0.351-μm light into a 600-μm-diameter volume. An optimally flat beam wavefront can achieve this pointing and focusing accuracy. The control system corrects wavefront aberrations by performing closed-loop compensation during laser alignment to correct for gas density variations. Static compensation of flashlamp-induced thermal distortion is established just prior to the laser shot. The control system compensates each laser beam at 10 Hz by measuring the wavefront with a 77-lenslet Hartmann sensor and applying corrections with a 39-actuator deformable mirror. The distributed architecture utilizes SPARC AXi computers running Solaris to perform real-time image processing of sensor data and PowerPC-based computers running VxWorks to compute mirror commands. A single pair of SPARC and PowerPC processors accomplishes wavefront control for a group of eight beams. The software design uses proven adaptive optic control algorithms that are implemented in a multi-tasking environment to economically control the beam wavefronts in parallel. Prototype tests have achieved a closed-loop residual error of 0.03 waves rms.

## 1 INTRODUCTION

A primary requirement for NIF is that each beam shall deliver its design energy into a 600-μm hohlraum ICF target. The total design energy for 192 beams is 1.8 megajoules. A goal for the system is that 50% of the design energy should be contained within a 100-μm focal spot at the target plane. This is about twice the diffraction-limited 80% energy spot size.

To meet the spot size requirement and goal, NIF subsystems are designed to limit wavefront aberrations. Optics have stringent specifications for rms surface gradient, power spectral density and surface roughness. Stringent specifications are also maintained for optical component mounting. NIF systems are designed to mitigate the effects of temperature and humidity variations and vibrations. An active alignment system is employed to point the beams accurately into the target. Even with these efforts to control wavefront aberrations, the spot size requirement and goal could not be met without a wavefront control system.

## 2 WAVEFRONT SYSTEM DESIGN

A block diagram of the NIF main laser optical system is shown in Figure 1, with the NIF Wavefront Control System components highlighted. The path of the NIF beam is first summarized and then the wavefront control functions applied to the NIF beam (or the wavefront control beam surrogate) are described.

The NIF preamplifier beam (1.053-μm wavelength, called 1ω) enters the main laser chain near the focus of the transport spatial filters (TSF), directed away from the target. The beam exits the filter collimated and passes through the booster amplifier heading toward the laser main amplifier cavity. A Pockels cell is set to allow the beam to enter the cavity, where it makes four passes through the main amplifier before the Pockels cell is switched to allow the beam to exit. The beam then exits the cavity, passes through the booster amplifier and the TSF, and heads towards the target chamber. The beam is frequency-converted to 351-nm wavelength, (called 3ω) at the target chamber.

Wavefront control functions are implemented as follows. Between shots, a continuous wave probe beam is co-aligned with the NIF beam prior to injection into the main laser. The probe beam follows the NIF beam path. A 39-actuator large-aperture deformable mirror (DM) operates at the far end of the laser cavity where the beam bounces twice. This two-bounce configuration doubles the effective stroke of the DM. At the TSF output, a tilted sampling surface reflects a small fraction of the probe beam towards a pick-off mirror near the TSF focus that sends the sampled beam through relays to the output sensor. Within the output sensor, a 77-lenslet Hartmann sensor measures the wavefront. The Hartmann sensor's video output is read by a framegrabber in the Wavefront Control System's Hartmann front-end processor (FEP). The Hartmann processor FEP transmits measured wavefront data to the mirror control FEP, which calculates the surface displacements to be applied to the DM to correct the wavefront aberrations in the beam.

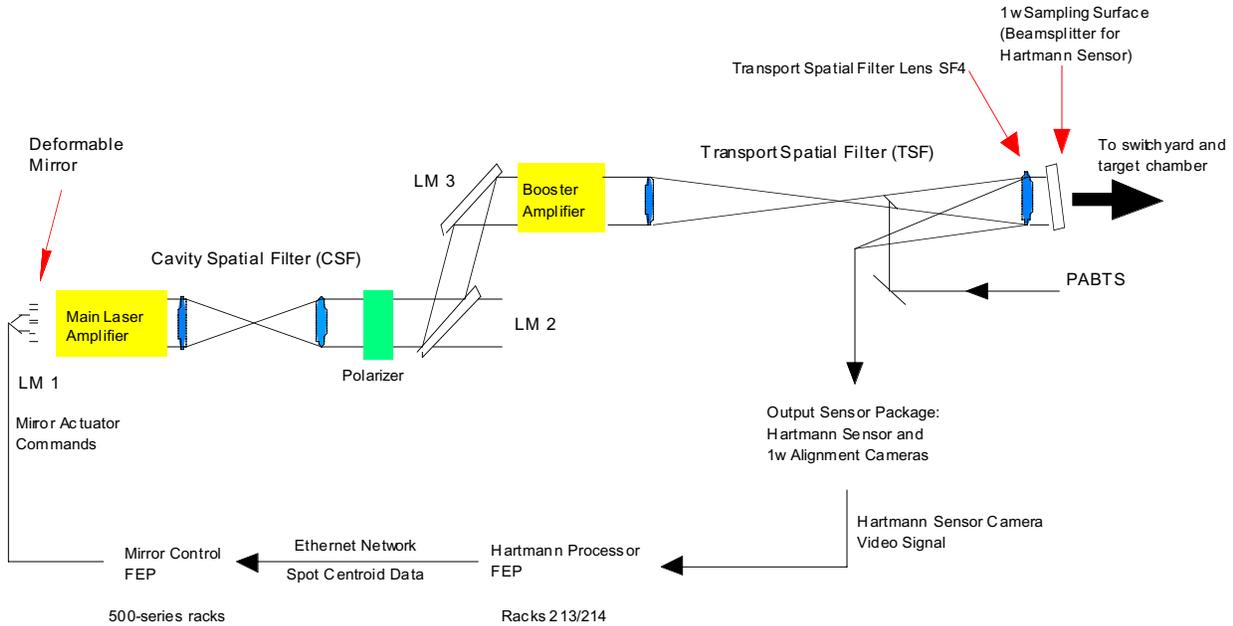

Figure 1: Block diagram of the NIF main laser optical system.

## 2.1 Wavefront System Requirements

There are several functions that the wavefront system should perform. Required functions are to: (1) control the 1ω output wavefront (including pointing) of each beam in the time immediately preceding a laser system shot, (2) apply compensation for previously measured pump-induced wavefront distortion, (3) measure beam output wavefront during a laser system shot, and (4) control the output wavefront (excluding pointing) during routine system operations between shots.

Other factors that influence the Wavefront control system requirements are expected system cost, and experience with Beamlet, the single-beam NIF prototype laser. The wavefront control system requirements are shown in Table 1. The initial design has met most requirements with the exception of closed-loop bandwidth. The bandwidth is processor-limited, and it is expected that by the time NIF is implemented, faster processors will be available to help meet this requirement without major changes to the software or design architecture.

## 3 SYSTEM CALIBRATION AND OPERATION

### 3.1 Control System Calibration

The wavefront measuring system is calibrated by inserting a wavefront reference fiber at the focal point of the TSF. Since the fiber light source is smaller than the TSF focal spot, the spot pattern at the Hartmann sensor when viewing the fiber reference beam is the same as the pattern the sensor would see when viewing the probe beam, if the beam and all upstream system components had diffraction-limited performance. The aberrations (imperfections in the separations of the lenslet array focal points) that are seen with the reference source inserted are due to the aberrations in the measuring system (sampling surface, relay optics, output sensor optics, and the sensor itself). By designing the control system to use the sensor focal spot image of the wavefront reference as the target wavefront to which the system wavefront is controlled, the system is being forced to generate, as closely as it is able, a perfect focal spot in the TSF. This also implies that all aberrations in all the optics beyond the TSF focus, including the TSF output lens, are uncorrected by the baseline wavefront control system.

Table 1: NIF Wavefront Control System Requirements

| Requirement | Value |
|---|---|
| Maximum residual low frequency spatial angle | 20 µradians |
| Maximum open-loop time before a shot | 1 second |
| Minimum closed-loop bandwidth | 1 Hz (0.5 Hz demonstrated) |
| Compensation range for simple curvature (double-pass reflected wavefront) | 15 waves at 1ω |
| Order of aberrations corrected | ≤4$^{th}$ order |
| Measurement accuracy at 1ω | 0.1 waves |
| Lenslet spacing | ≤1/2 of demagnified actuator spacing |

Next, the wavefront reference source is replaced by the probe beam, and the wavefront control system is calibrated by an online procedure. Each of the 39 actuators is individually poked and pulled relative to the best-flat starting point. The offset for each Hartmann spot is thus related to the displacement of each actuator. From this information, a gain matrix relating actuator movement to Hartmann sensor focal spot movement is derived.

*3.2 Nominal Closed-Loop Operation*

Once the calibrations are complete, the loop is closed wherein the measured Hartmann offsets from the reference positions are multiplied by the gain matrix, yielding the actuator offsets to control the mirror to flat (with appropriate loop gain for stability). This is the configuration used during pre-shot alignment. After alignment is complete and the shot sequence has begun, an additional Hartmann offset file is subtracted from the wavefront sensor data prior to being applied to the gain matrix. These additional offsets represent the uncorrected prompt pump-induced wavefront aberrations measured from a previous shot. By subtracting out these offsets, we are setting the wavefront to the conjugate of the expected prompt aberration of the upcoming shot. Thus, at shot time, the wavefront is flat.

The control system operates with a closed-loop bandwidth of up to 0.5 Hz. To achieve this, the wavefront sensor is read at a 10 Hz rate (with a goal of 30 Hz). The sensor is read in standard RS-170 video, which is captured by a DataCell Snapper 24 framegrabber. The digitized image is fed into a SPARCengine Axi computer that calculates centroids for all 77 lenslet spots. This information is sent via a dedicated ethernet line to the Motorola MVME 2306 controller that calculates the offsets of the spots from their reference positions. These offset data are input to the mirror control law, which calculates the required DM actuator displacements. Each Hartmann processor serves eight beams, and each mirror control processor serves four DMs. Each mirror control chassis contains two mirror control processors.

## 4 SOFTWARE DESIGN

The NIF wavefront control system software was designed using an object-oriented approach. The NIF facility is expected to be in operation for 30 years, and it is reasonable to expect to make software or hardware changes over that time. By using modular hardware and software architecture, the system maintainability is improved significantly.

The object structure of the wavefront control system essentially conforms to the generalized hardware interfaces (e.g. digital to analog conversion, network communication, image frame grabbing) and major sensing and control activities (spot tracking, display building, and control law processing) that take place. Abstract interface layers are used to minimize the effect on software of prospective hardware changes. This approach allows us to create as many beams per processor as the central processing unit hardware and memory will allow.

The real-time tasking design of the wavefront control software is basically interrupt-driven. The interrupt source is the video sync signal contained in the Hartmann sensor composite video captured by the framegrabber. Completion of this and each successive task (spot tracking, display building, spot data transmission) sends a signal to start the next successive task. Similarly, the mirror control processor waits for spot centroid data to arrive and processes it in a sensor task (which reads the data and subtracts the reference spot positions to produce the wavefront error), then the mirror task is signaled to use the wavefront error to compute the mirror control command.

## 5 RESULTS

In tests with the LLNL wavefront interferometry lab, the wavefront control system has achieved 30 Hz operation for one beam and a residual error of 0.03 waves rms when controlling the wavefront to a reference measured from an optical reference flat.